\documentstyle[prl,aps]{revtex}
\twocolumn
\newcommand{\BEQ}{\begin{equation}}                                           
\newcommand{\EEQ}{\end{equation}}
\newcommand{\BEA}{\begin{eqnarray}}                              
\newcommand{\EEA}{\end{eqnarray}}

\renewcommand{\H}{{\cal H}}

\begin{document}
\draft
\title {Diluted Generalized Random Energy Model}
\author{ D.B. Saakian }
\address
{ Yerevan Physics Institute \\
Alikhanian Brothers St.2, Yerevan 375036, Armenia}
\maketitle
\begin{abstract}
We introduce a layered random spin model, 
equivalent to the Generalized Random Energy Model
(GREM).
In analogy with diluted spin systems, a diluted GREM (DGREM) is introduced.
It can be applied to calculate approximately thermodynamic properties
of spin glass models in low dimensions. For Edwards -Anderson model it gives correct critical
dimension and 5\% accuracy for ground state energy in 2d. 
\end{abstract}

Derrida's Random Energy Model (REM) [1] was introduced as an 
archetype spin glass model. In recent years it is 
becoming more and more popular. 
It has been applied in many fields of physics, biology and even in 
information theory [2,3].

The REM is a mean-field model, like the Sherrington-Kirkpatrick model,
but maximally simplified. But even after this simplification 
it appears to be a good approximation of some real systems.

The generalization of the REM (called Generalized Random Energy Model, GREM)
was introduced in ref.  [5]. It was used to solve approximately other 
spin glass systems [6,7]. Unfortunately, the accuracy to 
describe other SG systems was not much better than for the REM.
In this work we introduce a diluted spin model which 
thermodynamically resembles GREM (in the case of large coordination number it is exactly equivalent to GREM),
then construct some new model of energy configurations-DGREM. In some (practically important) cases our spin model 
is thermodynamically exactly equivalent to DGREM. There is another case of DGREM also, where we could not
choose equivalent spin hamiltonian. \\
Our new model is  closer to realistic systems with 
a finite coordination number, than GREM. 

Even the simpler diluted REM (DREM) [8,9] has proven to be 
 a good approximation for models in low dimensions (d=1,2,3). 
This  important fact was observed in [10], 
where by information-theoretical arguments
(mathematically leading to a DREM) a percolation threshold was found.

We can estimate the ground state energy of the Edwards-Anderson model,
using a DREM  with the same number $z$ of non-zero couplings [8,9].
It holds that on a hyper cubic lattice in $d$-dimensions $z=Nd$, where $N$ is the number of spins.
For another lattices one should replace d by the half of coordination number.

In the DREM one has N Ising spins interacting with each other in the $z$ (randomly chosen from all the possible 
$C_N^p=\frac{N\verb|!|}{p\verb|!| (N-p)\verb|!|}$)
$p$-plets of Ising spins and quenched random couplings $\tau_{i_1,\cdots,i_p}$ having values $\pm 1$.

 The Hamiltonian reads
\begin{equation}
\label{l1}
\H=-\sum_{(1\le i_1,\cdots<i_p\le N)=1}^z\tau_{i_1,\cdots,i_p}
\sigma_{i_1}\cdots\sigma_{i_p}
\end{equation}

At high temperatures the system is in the para magnetic phase.
There it has a free energy
\begin{equation}
\label{l2}
\frac{F}{N}=-dT\ln \cosh \beta-T\ln 2
\end{equation}
where $\beta=\frac{1}{T}$
Below the critical temperature $T_c={1}/{\beta_c}$ the system freezes
in a  spin-glass phase with internal energy 
\begin{equation}
\label{l3}
\frac{U}{N}= -d\tanh \beta_c
\end{equation}
and vanishing entropy
$$S=0$$
 Here $\tanh\beta_c=f(d)$ involves a 
 function $f(x)$ defined by the implicit equation
\begin{equation}
\label{l4}
\frac{1}{2}(1+f)\ln (1+f)+\frac{1}{2}(1-f)\ln(1-f)]
=\frac{\ln 2}{x}
\end{equation}

Since the entropy is zero, we obtain for the 
ground state energy of the Edwards-Anderson model 
on a hyper cubic lattice in $d$ dimensions
\begin{equation}
\label{l5}
-\frac{E}{N}=f(d)d
\end{equation}
In 2d Esq. (5) gives $E\approx - 1.5599$, 
which is close to result of Monte-Carlo simulation for the
case of random $\pm 1$ couplings [11]
$\frac{E}{N}=-1.4015\pm 0.0008$.
 This estimate by formula (3) was done by Derrida in his original
work [1], long before the introduction of the DREM in reference [8].

Let us now construct some spin model, which has properties like the GREM. It is very important 
to have spin representation for GREM (for example- to construct temporal dynamics).\\
We consider a stacked system consisting of $M$ planes
with spin $\sigma_i^k$ ordered along a ``vertical'' axis.
In plane (layer) $k$ there are $N_k$ spins.
The Hamiltonian is a sum of $M-1$ terms $H_{k,k+1}$ and one  $H_M$. The last 
is as in eq. (1), with $N\to N_M,z\to z_M$ 
and all the interacting spins are from the 
$M$-th layer. For the $1<k<M-1$ term $H_{k,k+1}$ describes interaction between spins of neibor layers $k$ and $k+1$
through $z_k$ p-plets of spins. In this case in any p-plet  $\frac{p}{2}$  
 spins  are chosen from the layer $k$ and $\frac{p}{2}$ from the layer $(k-1)$. Our $z_k$ p-plets are chosen randomly 
 from the all possible $[\frac{N\verb|!|}{\frac{p}{2}\verb|!| (N-\frac{p}{2})\verb|!|}]^2$ ones.
There are no ``intra-plane''
interactions terms (besides the spins of layer M) 
but only ``inter-plane'' interactions.  
So spins in the layer $1<k<M$ interact with spins from the layers $k\pm1$, the first 
layer interacts with the spins of second layer.
We thus have the Hamiltonian
\begin{equation}
\label{l6}
\H=
-\sum_{(1\le i_1,\cdots<i_p\le N_M)}^{z_M}\tau_{i_1,\cdots,i_p}
\sigma_{i_1}^M\cdots\sigma_{i_p}^M
\end{equation}
$$-\sum_{k=1}^{M-1}\sum^{z_k}_{(1\le i_1,\cdots<i_\frac{p}{2}\le N_{k-1},1\le j_1\cdots <j_{\frac{p}{2}}\le N_k)}$$
$$\tau_{i_1,\cdots,i_p}\sigma_{i_1}^{k-1}\cdots\sigma_{i_\frac{p}{2}}^{k-1}\sigma_{j_1}^{k}\cdots 
\sigma_{j_\frac{p}{2}}^k$$
Till now we considered spin model with some interaction among the spins. Let us now introduce some
GREM like model. We consider some $M$ level hierarchic tree. At first level there are $2^N_1$ branches. At second level
every old branch fractures to $2^{N_2}$ new ones and so on. At the level $M$ there are 
 $2^N$ branches, where
\begin{equation}
\label{ll7}
N=\sum_{i=1,M}^M=N_i
\end{equation}
Energy configurations of our system are located on the ends of $M$-th level branches.
On  every branch of level $i$ there
 are distributed  random variables $\epsilon_i$ with the distribution
\begin{equation}
\label{ll7}
\rho_0(\epsilon_i,z_i)=\frac{1}{2\pi i}
\int_{-i\infty}^{i\infty}{\rm d}k \exp[-k\epsilon_i
+z_i\ln\cosh k]
\end{equation}
This is a distribution for a sum of $z_k$ random $\pm 1$ variables. So $z_k$ resembles number
of couplings in our diluted spin models. 
With any energy configuration are connected M branches. We define configuration energy as a sum
(along the path on the tree, connected with chosen energy configuration) of these
 $M$ variables 
 $\epsilon_i$.
 We see a usual picture of GREM, where random variable are distributed according to
(7) instead of normal partition.\\

We can consider the case of large $M$ with smooth distribution of 
$z_k$ and $N_k$. In this case we can introduce continues variable $v=\frac{k}{M}$
 between
$0$ and $1$, labeling the level of planes
and define distributions 

\begin{eqnarray}
\label{l7}
 z_k\equiv {\rm d}z=z{\rm d}v,\qquad  N_k\equiv {\rm d}N=n'(v){\rm d}v
\qquad{\rm d}{v}=\frac{1}{M}
\end{eqnarray}
where $n(v)$ is a given function (entropy in bits-s). The variable $v$ ( $0<v<1$) 
parameterizes the level of the hierarchical tree and $z$ is a parameter (for our spin system
z is a total number of couplings and parameter $v$ labeling the level of planes). \\ 
Of course, our function $n(v)$ should be monotonic. Total number of energy configurations
is $2^{n(0)}$ and 
\begin{equation}
\label{l8}
 n(0)=N
\end{equation}
We have, that $2^N$ energy levels $E$ of our hierarchic model are distributed by partition
\begin{equation}
\label{l9}
\rho(E)=\rho_0(E,z)
\end{equation}
If two configurations (in our GREM like model) meet at 
level of hierarchy $v$, they have  $zv$ common random variables.
The energy difference between two configurations is related 
to $z(1-v)$ non-common random variables. Therefore the 
 distribution function of two energies $E_1,E_2$ reads
\begin{equation}
\label{l10}
\rho_2 (E_1-E_2)=\rho_0((E_1-E_2),2 z (1-v))\exp (\ln 2 n(v))
\end{equation}
At high temperatures our system is in the para magnetic phase.
The free energy is given by eq. (2). 
When we decrease the temperature,  two situations are possible:
First, $\frac{dz}{dN}\equiv \frac{z}{n'(v)}$ decreases monotonically 
with  $v$; second,  it has a local maxim.\\
In the first case the system has no sharp phase transition but 
it freezes gradually. At temperature $T=\frac{1}{\beta}$ all level with
$0\le v \le v_f(T)$ are frozen; they are in the spin glass phase.
The levels with $v_f<v\le 1$ are in the paramagnetic phase.
$v_f$  is defined as the solution $v_f=v$ of equation
\begin{equation}
\label{l11}
\tanh\beta=f(\frac{z}{n'(v)})
\end{equation}
Using this relationship between $\beta$ and $v$ we can later use functions
$v(\beta)$ and $\beta(v)$.
For every finite $\beta$ the value of $v(\beta)$ lies between zero and unity.
When $T\to \infty$, $v(\beta)\to 0$ and when $T\to 0$,$v(\beta)\to v_0>0$.
So even in this limit some part of spins stay in their para magnetic phase.
Let us point out that this partial freezing only is possible 
in the Diluted GREM, and not in the original GREM.
 For the free energy we obtain ( there is no factor $N$ in it):
\begin{eqnarray}
\label{l12}
-\beta F=&\,& z(1-v(\beta))\ln \cosh \beta+n(v(\beta))\ln 2
\nonumber\\&+&
z\beta\int_0^{v(\beta)}{\rm  d}v_1f(\frac{z}{n'(v_1)}) 
\end{eqnarray} 
 The first two terms in the right hand side describes the para magnetic 
fraction of free energy ($n(v(\beta))\ln 2$ just is the entropy), while the last  one desrcibe the fraction
of spins frozen in a glassy configuration (it resembles equation (3) with $\frac{z}{f'(v_1)}$ instead of
d).\\
In the second case (function $n'(v)$ is not monotonic) the system has a sharp first order
phase transition at a finite temperature $T_2$. Below $T_2$
freezing occurs drastically for all levels  $v<v_2$,
 where $ v_2\equiv v(\beta_2)$ is 
defined by the equation
\begin{equation}
\label{ll11}
n'(v_2)=N-n(v_2)
\end{equation}
We used the fact that $n(0)=N$.  
The transition temperature $T_2=\frac{1}{\beta_2}$ follows from
\begin{equation}
\label{ll12}
\tanh(\beta_2)=f(\frac{z}{n' (v_2)})
\end{equation}
For temperatures $T<T_2$ the free energy reads
\begin{eqnarray}
\label{l13} 
-\frac{\beta F}{N}=&\,& z(1-v(\beta))\ln \cosh \beta+n(v(\beta))\ln 2 \nonumber\\
&+&
zv(\beta_2)\beta f(\frac{z}{n'(v_2)})
+z \beta\int_{v_2}^{v(\beta)}{\rm d}v_1f(\frac{z}{n'(v_1)})
\end{eqnarray}
Protein folding probably corresponds to this case.
 To construct the spin 
Hamiltonian by means of chain of subsystems 
for this case is  still an open problem. \\
Let us now consider a possible approximation to the Edwards-Anderson model,
following the idea's presented in ref.  [6].
In the $d$-dimensional case our
$2^N$ energy levels $E$ are distributed according to
the  law 
\begin{equation}
\label{l14}
\rho(\epsilon)=\rho_0(\epsilon,Nd)
\end{equation}
 with $\rho_0$ defined in eq. (8)
Comparing with (11) one immediately notices that
this is exactly equivalent to a DGREM with the choice 
$E=\epsilon$,  $z=Nd$. 
Let us now consider the distribution
of $\epsilon_1-\epsilon_2$. Following the arguments presented in
reference [6], we find that 
\begin{equation}
\label{l15}
\rho_2 (\epsilon_1-\epsilon_2)=
\rho_0 (\frac{\epsilon_1-\epsilon_2}{2},Nd(1-y))\exp Ns(-dN(2y-1))
\end{equation}
where $s(E)$ is the entropy of a ferromagnetic system on the same lattice
with internal energy equal $E$.
Physical meaning of  the last equation is transparent [6]. If we consider partition 
of two spin configurations, then one of them could be chosen (due to gauge
invariance) as ordered configuration with $\sigma_i=1$ for all spins.
In this case energy of first configuration equals to sum of $dN$ random $\pm$ couplings
and energy of second - sum of the $Ndy$  random variables with the same sign, and
$Nd(1-y)$ with the opposite sign. Distribution of second configurations by $y$ just equals
to number of configurations with energy $E=-(2y-1)Nd$ in ferromagnetic model.
 For the last in thermodynamic limit
we have expression $\exp Ns(-dN(2y-1)$.   
Unfortunately it differs from eq. (12). So it is impossible to choose
a DGREM with the given correlation functions for $E$ and $E_1-E_2$. We can 
consider an approximation. One choice is to take the exact expression for 
  $\rho (\epsilon)$ and make an approximation
 for $\rho (\epsilon_1-\epsilon_2)$.
\begin{equation}
\label{l16}
z=Nd,\quad \epsilon=E,\quad v=2y-1,\quad n(v)=\frac{Ns(-vdN)}{\ln 2}
\end{equation}
We see, that variable $v$, givimg distribution for the number of couplings among 
the hierarchy corresponds to energy per bond in ferromagnetic model.
We can remember from definition of temperature 
\begin{equation}
\label{l17}
\frac{{\rm d} s}{{\rm d } E}=\frac{1}{\tau}\equiv \beta
\end{equation}

At given   $\tilde \beta_1$ we can define corresponding $v_1$
 as minus energy per bond for  ferromagnetic
 model at temperature $\frac{1}{\tilde \beta_1}$.
\begin{equation}
\label{l18}
v_1=-\frac{E(\tilde\beta_1)}{Nd}
\end{equation} 

We obtain for the free energy
\begin{equation}
\label{l19}
-\frac{\beta F}{Nd}=(1-v(\beta))\ln \cosh \beta+s(v(\beta))+\beta\int_{0}^{v(\beta)}{\rm d}v_1
f(\frac{\ln 2}{\tilde\beta})
\end{equation}
Integrating by parts in the last term we get
\begin{eqnarray}
-\frac{\beta F}{Nd}&=&(1-v(\beta))\ln \cosh \beta+s(v(\beta))-\beta \int_{0}^{\tilde \beta}{\rm d}
\tilde \beta_1 \frac{2v_1(\tilde\beta_1)}{\ln\frac{1+y}{1-y}}\nonumber\\&+&v(\beta)\beta y(\tilde\beta) 
\end{eqnarray}

where $y$ as a function of $\tilde\beta_1$  is defined from the equation
\begin{equation}
\label{l120}
y=f(\frac{\ln2}{\tilde \beta_1 })
\end{equation}
function $v(\beta)$ is defined from (13), and $v_1(\tilde\beta_1)$ is minus energy per bond in ferromagnetic model
at temperature $\tilde\beta_1$.
In the  equation  (24) value of the $\tilde \beta$ is connected with the given $\beta$ via 
the equation
\begin{equation}
\label{l121}
\tanh (\beta)=f(\frac{\ln2}{\tilde \beta })
\end{equation}
In the limit of zero temperature {\bf this reduces to}
\begin{equation}
\label{ll22}
-\frac{\beta F}{ N}=d[1-\frac{2}{d}\int_{0}^{\ln 2}\frac{{\rm d}\tilde \beta_1 
E(\tilde \beta_1 )}{\ln\frac{1+y(\tilde\beta_1)}{1-y(\tilde\beta_1)}}]
\end{equation}
He $E(\tilde \beta)=|U|$ is minus the energy in the ferromagnetic model, 
$y(\tilde \beta)$ is defined from (25) and function $f(x)$
from the (4).
If the ferromagnetic system has a phase transition at  
$\tilde \beta_c$, then the  DGREM has one at $\beta_c$
{\bf determined by}
\begin{equation}
\label{l23}
\beta_c=\frac{1}{2}\ln\frac{1+ f(\frac{\ln2}{\tilde \beta_c})}{1-f(\frac{\ln2}{\tilde\beta_c})}
\end{equation}
This equation has solution only if
\begin{equation}
\label{l23}
\ln 2\> /\tilde \beta_c>1
\end{equation}
As for 2d Ising model  $\ln 2 <\tilde \beta_c$ , it follows that for the 2d
Edwards-Anderson model
 there {\bf does not occur a phase transition
at a finite temperature.}
is not any phase transition at finite temperatures.
{\bf A} calculation of {\bf the ground state} energy in this
case using (27) gives $E=- 1.4763$. 
 More generally,
for the case of other models
one can use numerical data for {\bf the} ferromagnetic system to
calculate {\bf the}
ground state energy of {\bf the} corresponding SG model on the
same lattice.\\
In the case of 3d Ising model we have for transition point $\beta_c=0.2216$.  In this case 
inequality (29) is satisfied. So our method gives as critical dimension for EA model $d=3$.\\
In conclusion, we have suggested a {\bf a diluted version of
the generalized random energy model } and applied to find 
a simple  approximation to the ground
state energy of disordered systems and definition of critical dimensions. 
\section*  {\it Acknowledgments}
D.B. Saakian thanks C.K. Hu for a possibility to
work in nice conditions of Academica Sinica (where this work was 
begun), P. Grassberger and H. Rieger for invitation to Juelich and discussions,
D. Bolle for invitation to Leuven and introduction to NN and especially to
P. Rujan for invitation to Oldenburg and very fruitful discussions.
The authors thank T.M. Nieuwenhuizen for a critical reading of the manuscript.

\end{document}